# Nonlinear Spectral Fusion Super-Resolution Fluorescence Microscopy based on Progressively Saturated Upconversion Nanoparticles

Yongtao Liu[1], Tianxiao Wu[1], Xiao Zhou[1], Fan Wang[2]

[1] *School of Electronic and Optical Engineering, Nanjing University of Science and Technology, Nanjing 210094, Jiangsu, China*

[2] *School of Physics, Beihang University, Beijing, 100191, China*

Email:Yongtao.Liu@njust.edu.cn;

**Abstract:** Single-beam scanning microscopy (SBSM) is one of the most robust strategies for commercial optical systems. Although structured illumination combined with Fourier-domain spatial spectrum fusion can enhance SBSM resolution beyond the diffraction limit, a sophisticated detection system is still required to optimize both effective resolution and signal-to-noise ratio.Here, we report that the diverse nonlinear responses of upconversion nanoparticles can unlock a new mode of Computational Progressively Emission Saturated Nanoscopy (CPSN), which employs a single doughnut-shaped excitation beam assisted by deep learning to simplify conventional microscopy. By modulating the excitation power, the smooth transition of the point spread function (PSF) from doughnut-shaped to Gaussian can be achieved, allowing for accessing different spatial frequency components of the sample. Then, in order to enhance time resolution, the doughnut-shaped beam at low power and the saturated Gaussian-like image were predicted by the doughnut-shaped beam at low saturation threshold based on the power dependence curve. Furthermore, a deep recursive residual network (DRRN) is employed to fusion these progressively complementary spatial frequency information into a final super-resolved image that encompasses the full frequency wwinformation. This approach can achieve high-quality super-resolution imaging with a spatial resolution of 33 nm, corresponding to 1/29th of the excitation wavelength, 55 dB of SNR ratio contracted to 7 dB in Gaussian imaging and applicable to any wavelength. The unique combination of nonlinear saturation and deep learning computational reconstruction could open a new avenue for simplifying the optical system and enhancing imaging quality in single-beam super-resolution nanoscopy.

## Introduction

Confocal Laser Scanning Microscopy (CLSM) is based on the fact that the pinhole has a powerful filtering and layer cutting effect while detecting the signal. The out-of-focus signal is effectively suppressed, and the imaging result with a higher signal-to-noise ratio than the wide-field microscope is obtained. By taking advantage of the system simplicity due to its single beam, CLSM is suitable for commercial applications. To solve the problem of resolution, the point scanning super-resolution system started with the CLSM and developed many improved point- scanning microscopy technologies.[1]Recent endeavors have sought to combine the system simplicity of CLSM and the super-resolution characteristics of STED. One promising avenue involves leveraging the nonlinear response characteristics to achieve super-resolution using a single doughnut-shaped excitation beam, which simplifies the optical configuration and reduces excitation power, minimizing phototoxicity. [6–11] Previous studies have employed upconversion nanoparticles to develop a super-resolution microscopy technique that utilizes heterochromatic Fourier-domain fusion, which expands the spatial information and enhances resolution by scanning with a single doughnut-shaped beam. By exploring the nonlinear properties of fluorescent probes, new modalities including approaches like Fourier fusion [12], saturated structured illumination [13] and multiplexed PSF[14], have demonstrated the capability to extract high-spatial-frequency information for enhanced super-resolution imaging. However, as a prerequisite for extracting more comprehensive spatial frequency information to enhance overall image resolution and signal-to-noise ratio, these advances still require specialized fluorescence probe with multiple wavelengths to generate multiplexed PSF images. Additionally, a key condition is to find the appropriate threshold, which balances imaging resolution and signal-to-noise ratio. Therefore, the complex detection pathways, complicated fluorescence probe, and the need for high imaging quality remain key limitations for the widespread application of single-beam scanning microscopy.

Lanthanide-doped UCNPs overcome the high-spatial frequency limitations of conventional fluorophores by converting two low-energy near-infrared photons into a single high-energy photon at shorter wavelengths. [15,16] Their multiple intermediate energy levels facilitate upconversion emission via stepwise energy transfer, [17,18] resulting in a nonlinear response between excitation power and electron generation rate. [19–23] Recent research has actively explored this nonlinearity, leading to innovative approaches for breaking the diffraction limit by these unique optical properties of UCNPs. This includes techniques such as super-linear excitation-emission (SEE)[24], heterochromatic nonlinear fusion [12], photon avalanche (PA)[23], migrating photon avalanche (MPA)[20,21], and nonlinear upconverting stimulated emission depletion (U-STED) microscopy[25], which respectively leverage super-linear emission, a 26th order nonlinear response in $Tm^{3+}$-doped UCNPs, 46th order nonlinearity via ion avalanching in multilayer core/shell nanostructures, and dynamic cross-relaxation (CR) energy transfer, to achieve enhanced spatial resolution and surpass the diffraction limit.

Here, we present a high-order nonlinear Computational Progressively Emission Saturated Nanoscopy (CPSN) imaging technique based on deep learning. By employing a single doughnut-shaped beam to obtain an image at low saturation threshold, we developed a computational evolution method based on material prior characterization to calculate a distinct optical transfer function (OTF) with unique frequency characteristics. With the assistance of Deep Recursive Residual Network (DRRN) [26], we effectively capture the unique frequency characteristics encoded within a sequence of power-dependent PSFs and fuse Fourier components from each computational images based on power-dependent curve, ultimately achieving super-resolution imaging. This approach maximizes information utilization, eliminating the need for manual threshold selection and effectively addressing information mismatch. This approach significantly streamlines data processing, enhances SNR, and effectively minimizes imaging artifacts, ultimately leading to improved image quality.

**Results**

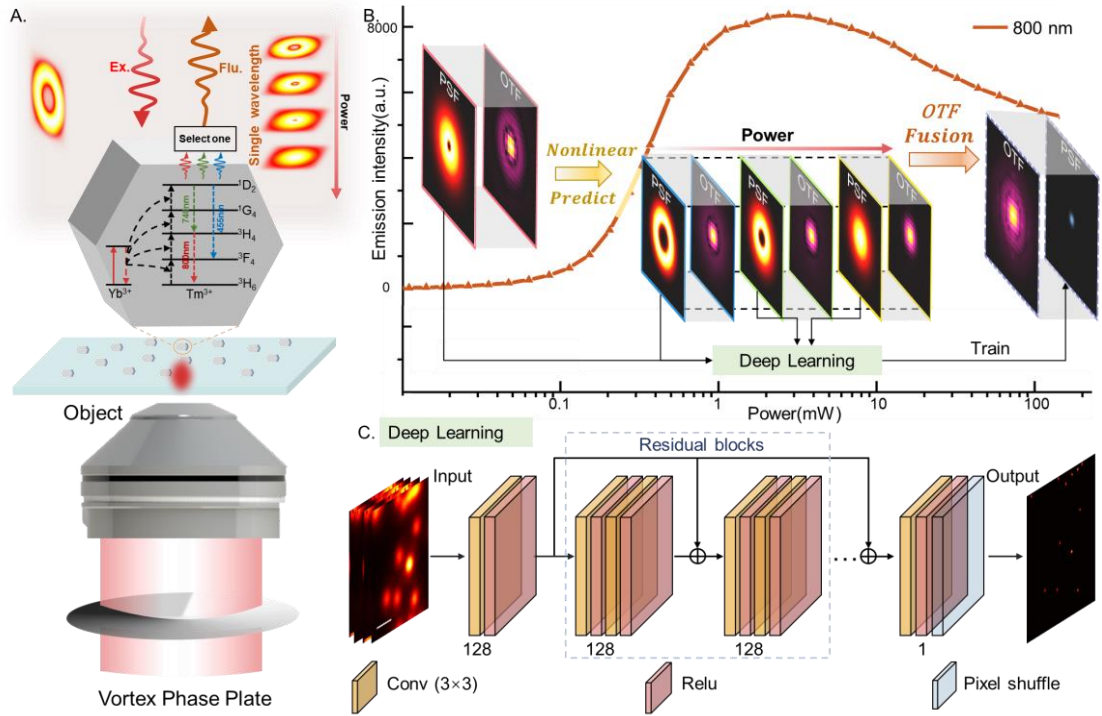

Figure 1. Concept of multi-order nonlinear upconversion super-resolution imaging based on deep learning. (A) Simplified energy level of nanoparticles $NaYF_4$:40% $Yb^{3+}$, 4% $Tm^{3+}$ under 980 nm excitation. The multi-photon near-infrared (NIR) upconversion emissions mainly from the two-photon excited state (740 nm, $^1D_2 \rightarrow {^3H_4}$ and 800 nm, $^3H_4 \rightarrow {^3H_6}$) and four-photon excited state (455 nm, $^1D_2 \rightarrow {^3F_4}$). The simulation results of the power-dependent PSF patterns of attribute emission of a single UCNP under the 980 nm doughnut-shaped beam. A vortex phase plate (VPP) is used to generate a doughnut excitation beam at the sample plane. (Details of negative confocal setup shown in Supplementary Information Note 1). (B) The power-dependent saturation intensity curves of the 800nm emissions. The inset is the PSFs and OTFs of doughnut PSF, predicted PSFs and OTFs under different power, OTF and PSF of fusion results and the simplified process of Deep Learning. (C)Deep-Fusion training pipeline. Based on the deep recursive residual network, Deep-Fusion takes four captured images respectively from the Gaussian PSF and doughnut PSF under different power as input to the generator, and yields a super-resolution result. Scale bar is 1 μm.

We employ upconversion nanoparticles (UCNPs) as fluorescent probes to demonstrate our strategy because their energy levels exhibit differential nonlinear responses that enhance resolution. We choose the particles of $NaYF_4$ doped with 4% $Tm^{3+}$ and 40% $Yb^{3+}$(Supplementary Information Note 2) . The sensitizer ion $Yb^{3+}$ directly absorb 980 nm near-infrared photons and transfer the energy to the activator ion $Tm^{3+}$ with a stepwise energy transferring process, leading to the emission of fluorescence at 740 nm ($^1D_2 \rightarrow {^3H_4}$), 800 nm ($^3H_4 \rightarrow {^3H_6}$) and 455 nm ($^1D_2 \rightarrow {^3F_4}$) (Figure 1A). To explore this nonlinearity, we employed a custom-built single-beam super-resolution imaging system, analogous to STED microscopy but based on negative confocal microscopy (Figure 1A, SI Figure 1).To further verify the effective increase of spatial frequency information by UCNPs，we measured the power-dependent fluorescence emission curve of near-infrared fluorescence at 800 nm, which behaves a two-photon-like nonlinearity and forms the foundation of our super-resolution imaging approach (Figure 1B). We further investigated this nonlinear behavior by examining the fluorescence point spread function (PSF) at various excitation powers. With increasing excitation power, the 800nm fluorescence PSF exhibited a doughnut shape because of nonlinear saturation, eventually transitioning to a Gaussian-like distribution upon reaching the saturation threshold (Figure 1B inset). To further validate our hypothesis regarding the information content within these power-dependent PSFs, we transformed them into their corresponding optical transfer functions (OTFs) within the Fourier domain. As anticipated, the progressive doughnut-shaped PSFs exhibited an enrichment of high-spatial frequency information, while the Gaussian-like PSF provided the missing information in the intermediate/ low-frequency components (Figure 1B inset). To improve the imaging speed, we predict images at other powers based on the power dependent curve using a doughnut image at a low saturation threshold, including a doughnut image at lower power and a saturated Gaussian-like image. To get the resultant emission PSF effectively covered the wider spectrum range of both low- and high-frequency information, we perform frequency shifting mechanism which fuse the fluorescence OTFs with specific responses under different excitation powers in the Fourier domain (Figure 1B inset). Nevertheless, achieving this broad frequency coverage also introduced new challenges. The image artifacts are serious and the extensive deconvolution process was complex and time-consuming.

To address the inherent challenges of Fourier domain fusion algorithms, particularly the information mismatch arising from multi-image fusion, as well as issues like imaging artifacts, information loss, and processing complexity, we employ the Deep Recursive Residual Network (DRRN). This network is composed of 25 recursive residual blocks, each containing two convolutional layers with a ReLU activation function following the first. As shown in Figure 1C, we first constructed a dataset containing four images convolved with Gaussian and different doughnut PSFs with random particle points, in other words, these four scanning images of UCNPs under different powers are used as input to the network, which then go through the first convolution layer for feature extraction and fusion. Subsequently, the output is processed by the recursive residual blocks. This cascaded structure facilitates progressive feature enhancement, allowing the DRRN to effectively extract intricate image details. During the processing in each residual block, features were added to the initial input features to achieve residual learning, enabling "parameter sharing" and obtaining more convolution with fewer parameters. Finally, features refined by the residual blocks are merged through a concluding convolutional layer and upsampled to generate a high-resolution image. Additionally, this network is optimized using the Mean Squared Error (MSE) loss function, enabling it to generate super-resolution images exhibiting enhanced detail, richer textures, and superior resolution. This superior performance is further validated across multiple datasets, demonstrating the DRRN's advantages in image recovery quality and model efficiency, underscoring its potential for diverse image processing applications.

**Numerical Simulation**

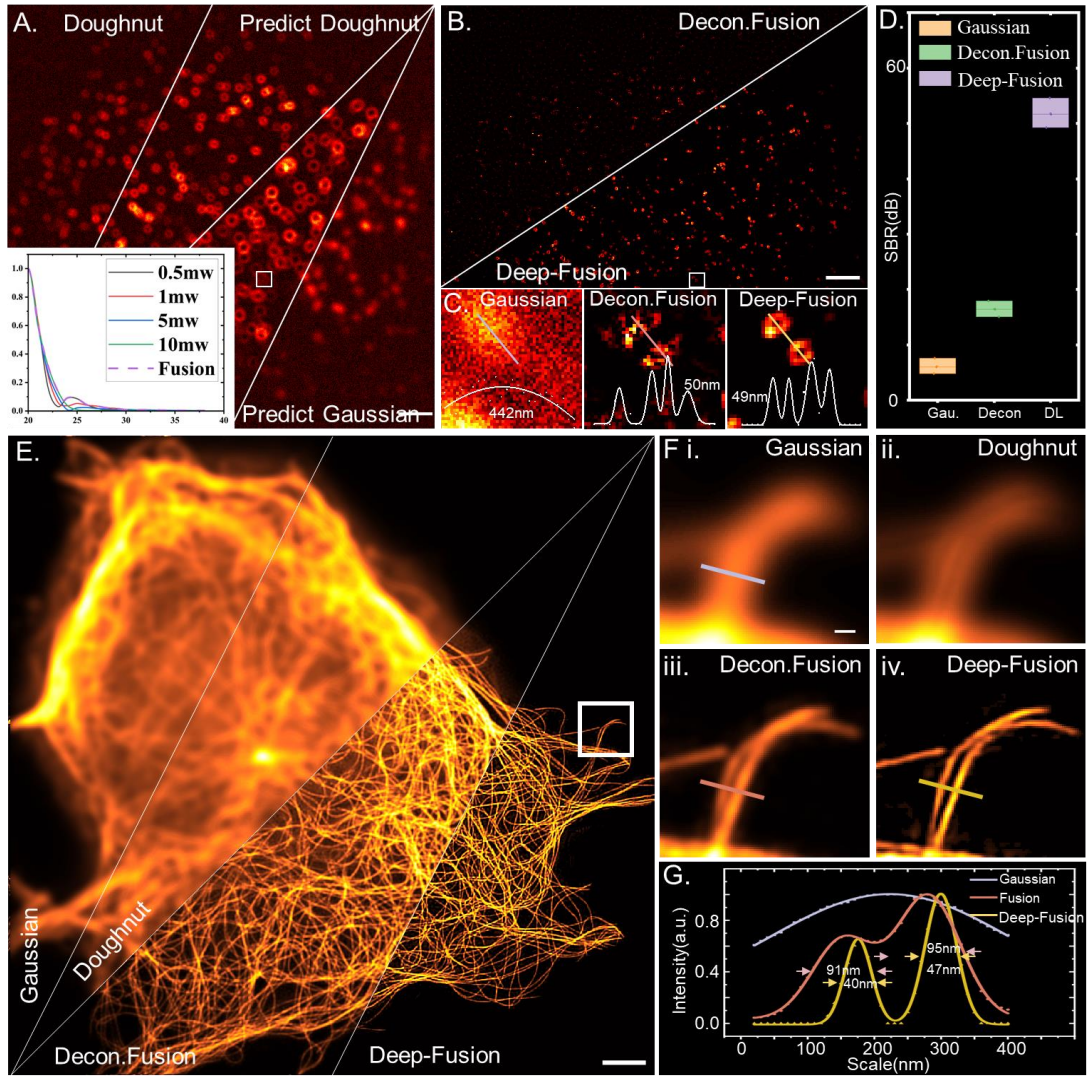

Figure 2. Theoretical simulation results of CCPs and microtube. (A) Simulation scan images of CCPs with doughnut-shaped beams under 5 mW accompanied by Poisson noise, predicted doughnut-shaped image at 0.5 mW and 1 mw, predicted Gaussian-like image at 10 mW based on the power dependent curve. (B) Comparison of super-resolution images obtained by Fourier domain fusion deconvolution (Decon.Fusion) and deep learning Fourier domain fusion algorithm (Deep-Fusion). (C)Magnified region of the white square, showing Gaussian(i), Decon.Fusion(ii), and Deep-Fusion(iii) magnified images. (D) Signal-to-Noise Ratio Analysis of predicted Gaussian images, Decon.Fusion, and Deep-Fusion. (E) Simulation scan images of microtubules with Gaussian and doughnut-shaped beams, as well as Fourier domain (Decon.Fusion) and deep learning images. (F) Magnified region of the white square, showing Gaussian(i), doughnut-shaped(ii), Fourier domain fusion deconvolution(iii), and Deep-Fusion magnified images(iv). (G) Intensity cross-sectional profiles along the line regions in panel F. Scale bar is 1 μm in (A), (B), 200 nm in (C), 2 μm in (E), 200 nm in (F).

To demonstrate the SNR enhancement achieved by our method, we conducted numerical simulations on Clathrin-Coated Pits (CCPs), comparing the performance of doughnut-shaped beams, predicted doughnut-shaped beams under lower power, predicted Gaussian-like emission under saturated excitation, Fourier domain fusion, and deep-learning enhanced Fourier domain fusion. As shown in Figure 2A, we simulated CCP scanning images under 5 mW power by convolving a ground truth CCPs image with doughnut PSFs and predict other images under different power including 0.5 mW, 1 mW and 10 mW. We then transformed these simulated images into their Fourier domain representations using a 2D fast Fourier transform. Next, we created a "segmented Fourier image" (Figure 2A inset, details see SI Figure 3) by segmenting and combining these Fourier domain images using a Fourier binary phase mask. The amplitude at a specific spatial frequency represents the achievable resolution, with cutoff frequencies (fcut1, fcut2, fcut3) defined by the intersections of the various doughnut OTFs and the Gaussian-like OTF. The fused OTF included the maximum amplitude envelope parts of the 10mw Gaussian-like emission (spatial frequency from 0 to $f_{cut1}$), 0.5 mw doughnut emission (spatial frequency from $f_{cut1}$ to $f_{cut2}$), 1mw doughnut emission (spatial frequency from $f_{cut2}$ to $f_{cut3}$), and 5mw doughnut emission (spatial frequency > $f_{cut3}$). The fused image was further processed by inverse Fourier transform followed by deconvolution to obtain the final super-resolution imaging result. Our imaging results (Figure 2B) highlighted the limitations of relying solely on Gaussian or doughnut-shaped PSFs for super-resolution.

Accurately determining the optimal frequency threshold points for fusing images with varying SNR is a significant challenge, as this process is prone to errors that can lead to information loss and mismatches, ultimately compromising the final resolution and introducing image artifacts and distortions. Moreover, it lacks generalizability and requires manual fine-tuning, adding to the complexity and time consumption, especially when considering the computationally intensive deconvolution process. To overcome these limitations, we propose our DRRN-based Deep-Fusion method, which directly learns to fuse the rich spatial frequency information contained within doughnut-shaped PSFs at various excitation powers. The recursive structure where the same residual blocks are repeatedly applied allows it to progressively capture and refine fine details across multiple iterations and enhances the network's ability to focus on intricate features. The residual blocks within the DRRN are designed to optimize residual learning between the input and output images, enhancing the network's sensitivity to edges and details, capturing richer image features, and improving robustness towards noise. As a result, the imaging result obtained by Deep-Fusion, as shown in Figure 2B, magnified region of the white square shows the enhanced resolution and SNR more clearly (Figure 2C). Along with the cross-section profile (Figure 2C inset), demonstrates that Deep-Fusion achieves 49 nm resolution. The box plot analysis of the SNR ratio further shows that Deep-Fusion improves the SNR ratio from 15 to 54 compared with the traditional method. Our method not only surpasses the Fourier domain fusion method in terms of imaging resolution but also significantly improves the SNR.

To further validate the reliability of our Deep-Fusion method, we conducted numerical simulations using microtubule patterns, as illustrated in Figure 2E. The magnified comparison images (Figure 2F), focusing on the regions marked by white squares, showcase the superior imaging quality achieved by Deep-Fusion (Figure 2F(iv)). In contrast, both Gaussian-like illumination (Figure 2F(i)) and the Fourier domain fusion method (Figure 2F(iii)), which incorporates high-frequency information from doughnut illumination (Figure 2F(ii)), exhibit limitations in resolving fine details. The corresponding line profiles (Figure 2G), fitted with Gaussian functions, reveal a resolution of 91 nm for the Fourier domain fusion method, while our Deep-Fusion method achieves a superior resolution of 40 nm. This enhanced performance stems from the DRRN's unique architecture. Its recursive and residual structure facilitates efficient feature extraction and fusion, while its optimized training strategy ensures robustness across diverse input image types. Consequently, our deep learning-based spectral fusion method not only effectively mitigates image distortions and significantly accelerates the reconstruction process, but also circumvents the challenges of accurately determining frequency threshold points for images with varying SNR, a process prone to errors leading to information loss, mismatches, compromised resolution, and image artifacts inherent to conventional Fourier domain fusion.

**Experiments**

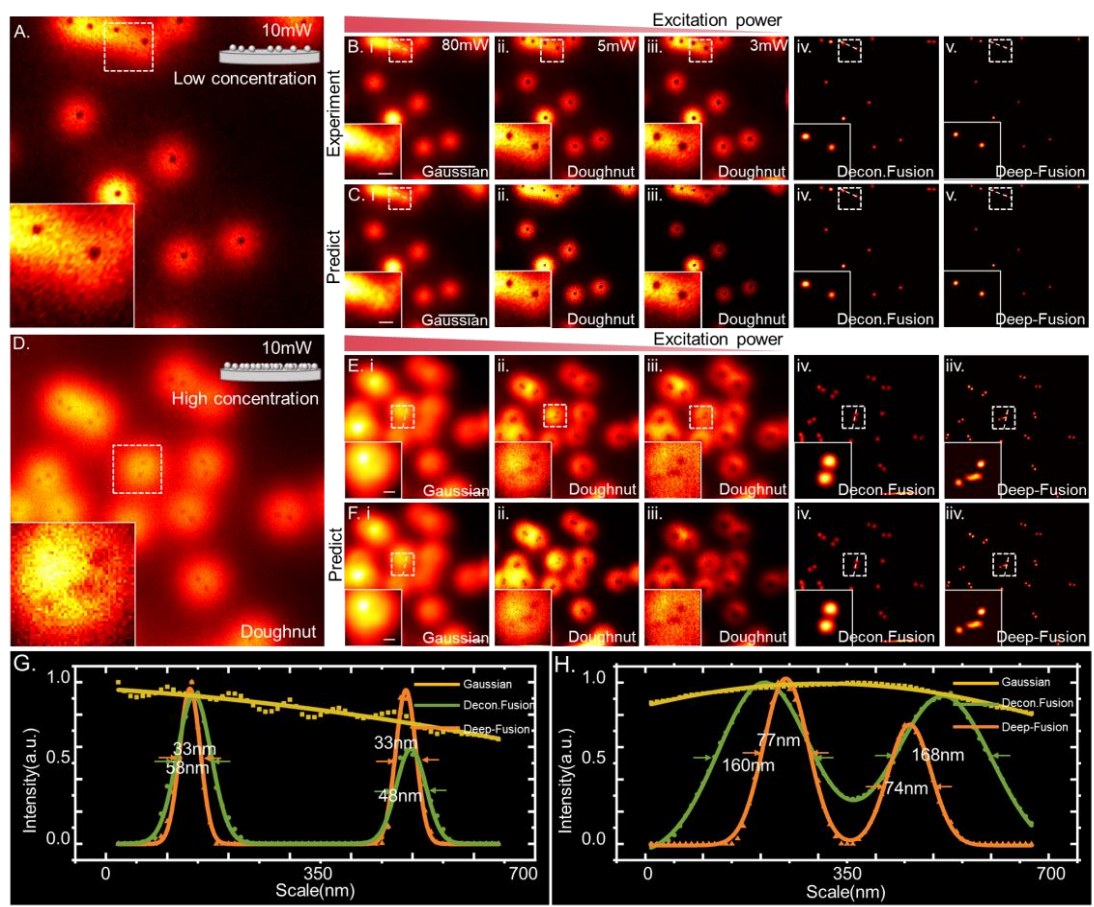

Figure 3. Experimental results of UCNPs in sub-diffraction volume. (A-B) The 800 nm emission band image of single UCNPs under a 980 nm doughnut beam with 10 mW, 80 mW, 5 mW, and 3 mW, respectively. The super-resolution imaging result by Fourier domain fusing the OTFs of (A-B(iii)) (Decon.Fusion). The output of the DRRN architecture (Deep-Fusion). (C)The predicted results of doughnut and Gaussian-like beams. The traditional Fourier domain results and Deep-Fusion results. (D-E) The 800 nm emission band image of dual-fluorescence nanorods under a 980 nm doughnut beam with 10 mW, 80 mW, 5 mW, and 3 mW, respectively. The super-resolution imaging result by Fourier domain fusing the OTFs of (D-E(iii)) (Decon.Fusion). The output of the DRRN architecture (Deep-Fusion). (F) The predicted results of doughnut and Gaussian-like beams. The traditional Fourier domain results and Deep-Fusion results of dual-fluorescence nanorods. Inset is the magnified region of the white square. (G) Line profiles of two nearby UCNPs from (A) to (C). (H) Line profiles of four nearby UCNPs from (D) to (F). Scale bar is 1 μm in (A-F), 200 nm inset.

Based on simulation results, we experimentally validated our Deep-Fusion method using a custom-built single-doughnut beam super-resolution microscopy system to scan UCNPs exhibiting nonlinear stepwise saturated excitation. Acquiring 800 nm fluorescence images under stepwise saturated doughnut scanning with a two-photon-like confocal system (Figures 3A-B). Leveraging the inherent nonlinear excitation characteristics of UCNPs, we observed a distinct transition from multi-order doughnut-shaped PSFs to a Gaussian-like PSF with increasing excitation power (Figures 3A-B). As expected, Gaussian-like imaging, limited by the diffraction limit, failed to resolve individual nanoparticles (Figure 3B(i)). In contrast, negative confocal imaging provided higher frequency information, enabling the resolution of single UCNP positions (Figure 3A-B(ii)& (iii)). In order to verify the prediction of the imaging quality of images at other powers based on the power dependence curve, we further compared the super-resolution results obtained from the experimental images and the predicted images (Figures 3C). We further enhanced resolution by applying a Fourier domain fusion algorithm, combining OTFs from Gaussian-like and negative confocal imaging, achieving a resolution of up to 48 nm (Figure 3B(iv)&3C(iv)). However, as predicted by our simulations, this method was susceptible to artifacts and resolution degradation in more complex samples. We then applied our Deep-Fusion model, trained on simulated datasets, to the acquired Gaussian-like and negative confocal images. The resulting super-resolved image (Figure 3B(v)&3C(v)) demonstrated the ability to resolve discrete nanoparticles at a distance of 33 nm (Figure 3G). To further validate our method's feasibility and robustness, we imaged dual-fluorescence nanorods with a fixed 210 nm spacing (Figure 3D-E). As shown in the magnified comparison images (Figure 3D-E inset), the Fourier domain fusion method struggled to accurately recover high-concentration sample particles, generating artifacts (Figure 3E(iv) inset). In contrast, Deep-Fusion effectively addressed these challenges, clearly resolving the paired fluorescence particles (Figure 3E(v) inset). The predicted results are similar to the experimental results(Figure 3F).The corresponding line profiles (Figure 3H) confirm a significant resolution enhancement to 74 nm for single UCNPs. These results highlight the power of our Deep-Fusion method, which combines the nonlinear saturation response of UCNPs with convolutional neural networks to achieve superior resolution and image interpretability without adding complexity to the optical setup.

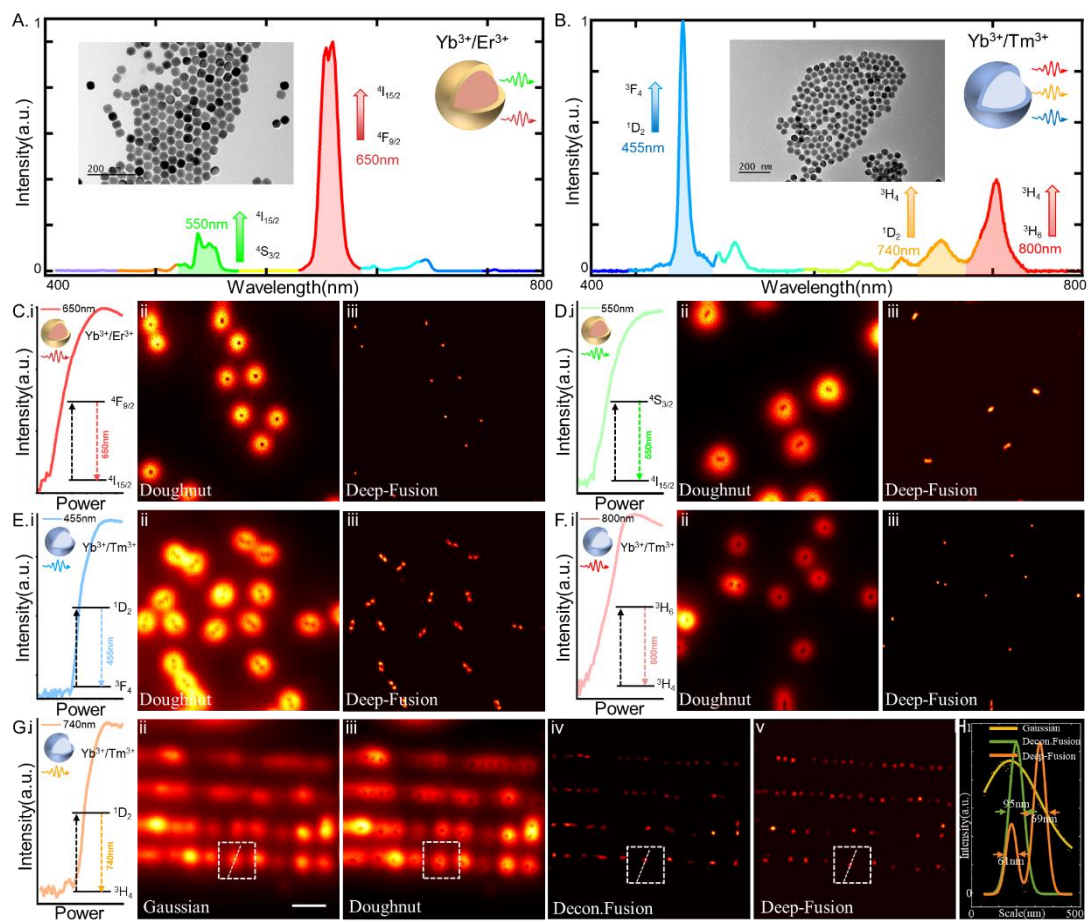

Figure 4. Experimental results of dispersed and dense samples in sub-diffraction volume of different wavelength. (A) Spectral

images of UCNPs co-doped with $Er^{3+}$ and $Yb^{3+}$ (NaYF$_4$: 2%$Er^{3+}$, 20%$Yb^{3+}$). (B) Spectral images of UCNPs co-doped with $Tm^{3+}$ and $Yb^{3+}$ (NaYF$_4$: 4%$Tm^{3+}$, 40%$Yb^{3+}$). TEM image of the UCNPs. (C-F) Power dependent curve of 650nm, 550nm, 455nm and 740nm fluorescence, negative confocal scanning image of UCNPs and the final Deep-Fusion results. (G) The 800 nm emission band image of dense UCNPs under a 980nm doughnut beam with 80mW(ii), 10mW(iii), the super-resolution imaging result by Fourier domain fusing the OTFs of (ii) and (iii) (Decon.Fusion) (iv) and the output result of the DRRN architecture(v), respectively. F) Line profiles of two nearby UCNPs in (G).

To further demonstrate the versatility of our method, we utilized UCNPs co-doped with $Tm^{3+}$ and $Yb^{3+}$ (NaYF$_4$: 4%$Tm^{3+}$, 40%$Yb^{3+}$) and UCNPs co-doped with $Er^{3+}$ and $Yb^{3+}$ (NaYF$_4$: 2%$Er^{3+}$, 20%$Yb^{3+}$). Figure 4A and B show the spectra of differently doped UCNPs. In order to demonstrate the particle uniformity of UCNPs, we also include TEM images of UCNPs (Figure 4A and B inset). Firstly, we perform single doughnut beam scanning, focusing on the 650 nm fluorescence channel and measuring the power dependent curve (Figure 4C(i)). Figure 4C(ii)&(iii) display the acquired images, showcasing both negative confocal imaging and Deep-Fusion results. To further verify the robustness of microscopy, we collect 550 nm fluorescence signal and obtain related power dependent curve (Figure 4D(i)). After processing the experimental image (Figure 4D(ii)) using our convolutional neural network, we observed a substantial resolution enhancement Figure 4D(iii).

To verify the effectiveness and resolution enhancement of our Computational Progressively Emission Saturated imaging technique, we imaged UCNP samples in attribute fluorescence wavelength. Firstly, we imaged dual-fluorescence nanorods under 455 nm fluorescence and get the power dependent curve of 455 nm. Based on the power dependent curve and with the help of deep learning, we obtain the final super resolution results (Figure 4E(iii)) from the single negative confocal experiment scanning image (Figure 4E(ii)). Furthermore, we image disperse UCNPs under 740nm fluorescence channel. As shown in the Figure 4F, our Deep-Fusion method clearly resolve all the particles.

Next, we assessed Deep-Fusion's performance on densely packed UCNPs under 800 nm fluorescence channel. Based on the power dependent curve of 800 nm fluorescence, we obtain different images and Gaussian-like imaging shows weakened pattern features due to high sample concentration. While Fourier domain fusion, incorporating high-frequency doughnut imaging information (Figure 4G(ii) & 4G(iv)), could distinguish some particles, it lost information for weaker particles and failed to resolve closely spaced particles. In contrast, Deep-Fusion clearly distinguished densely packed particles (Figure 4G(v)) and recovered weaker signals. The cross-section profile (Figure 4H) confirmed a significant resolution improvement from 95 nm with Fourier domain fusion to 61 nm with Deep-Fusion, highlighting its effectiveness and generalizability in dense samples. These experiments show that Deep-Fusion leverages the high-frequency information of the doughnut-shaped beam and the assistance of deep learning to achieve excellent resolution and robust performance at different sample densities. Most importantly, it can achieve good imaging quality in any fluorescence channel and is universally applicable to fluorescent probes.

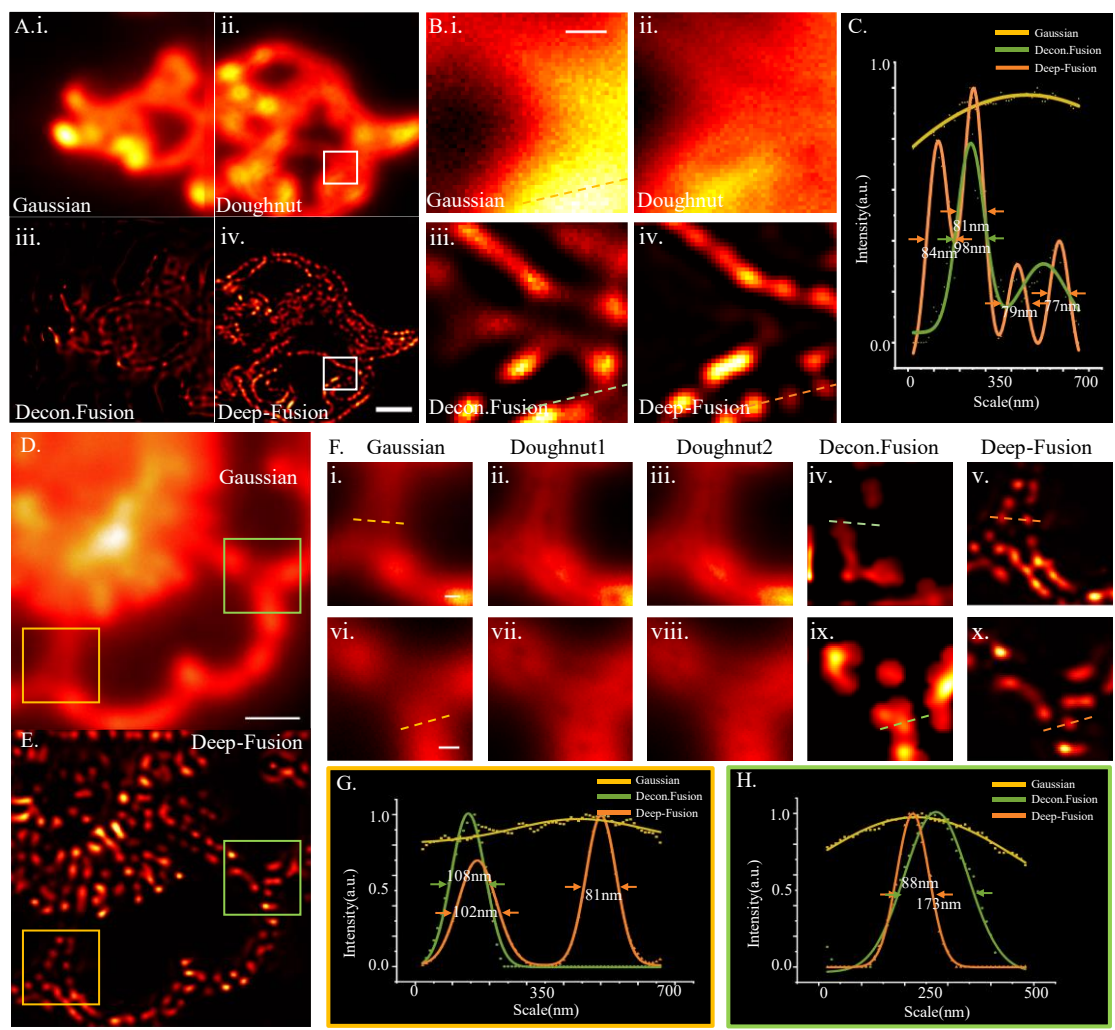

Figure 5. The overall enhanced image quality in super-resolution. A) The confocal, doughnut scanning image, Fourier domain fusion deconvoluted and Deep-Fusion images of the dragon pattern. B) The magnified region of the white square in A. C) Line profiles of four nearby UCNPs in B. D-E) The confocal and the Deep-Fusion images of a sunflower pattern. F) Magnified region of the green and yellow square of the image results of a sunflower pattern from different methods: i) and vi) confocal imaging; ii) and vii) doughnut imaging under 5mW; iii) and viii) doughnut imaging under 10mW; iv) and ix) Fourier domain fusion deconvoluted; v) and x) Deep-Fusion result; G-H) Corresponding cross-line profiles in yellow and green square, respectively. Scale bars: 1 μm in A, B, E and F; 200 nm in B and F.

To further demonstrate the versatility of our Computational Progressively Emission Saturated imaging technique, we applied it to resolve UCNPs assembled into various large-scale patterns, specifically focusing on a dragon and a sunflower pattern. For the dragon and sunflower pattern, we utilized a pre-fabricated substrate featuring a nanohole array meticulously crafted through electron-beam lithography (EBL). This nanohole array served as a template for UCNP deposition, resulting in a high-resolution dragon pattern composed of precisely positioned UCNPs. Gaussian-like imaging (Figure 5A(i) & 5B(i)) failed to resolve individual particles due to the diffraction limit. Doughnut beam scanning, however, provided high-frequency information, effectively indicating individual particle positions (Figure 5A(ii) & 5B(ii)). While the Fourier domain fusion method resolved some details (Figure 5A(iii)), it suffered from severe artifacts in dense regions (Figure 5B(iii)). In contrast, our DRRN-based Deep-Fusion method significantly improved SNR and clearly resolved particles even in dense areas (Figure 5A(iv) & 5B(iv)). Line profiles (Figure 5C) confirmed

Deep-Fusion's ability to resolve single particle points at a distance of 77 nm, surpassing the capabilities of the traditional Fourier domain fusion algorithm. To highlight Deep-Fusion's ability to compensate for missing spatial frequencies, we analyzed a sunflower pattern. Gaussian-like saturated imaging lacked effective high-frequency information (Figure 5D & 5F (i & vi)). Subsequent doughnut beam scans at two power levels demonstrated that as the excitation approached the saturation threshold, finer structural details emerged due to the inclusion of more high-frequency information (Figure 5F (ii & iii & vii & viii)). However, the Fourier domain fusion algorithm, even with information from three power levels, produced images with excessive background noise and significant information loss, failing to accurately reconstruct the sunflower pattern (Figure 5F(iv & ix)). Deep-Fusion, on the other hand, significantly improved image quality (Figure 5E & 5F (v & x)), achieving a resolution of up to 81 nm and 88 nm (Figure 5G & H).

In summary, we have presented Computational Progressively Emission Saturated Nanoscopy (CPSN), a novel approach for single-doughnut beam super-resolution imaging that harnesses the nonlinear responses of UCNPs and the power of deep learning. By modulating the excitation power of a single doughnut-shaped beam, we generate a spectrum of excitation states within the UCNPs, effectively encoding a range of spatial frequency information within their progressively saturated PSFs. We then employ a Deep Recursive Residual Network (DRRN) to fuse this progressively acquired spatial frequency information, generating a final super-resolved image that encompasses the full spectral content. This approach, benefiting from deep learning-driven spectral fusion, significantly improves SNR and mitigates artifacts, maximizing information extraction without the need for a separate depletion beam as in conventional STED. Consequently, CPSN offers several advantages: reduced system complexity, minimized phototoxicity, and enhanced image quality with improved resolution, minimal distortion, and reduced information loss. While this study demonstrates the potential of DRRN for spectral fusion, obtaining optimal results with limited training data remains a challenge. Future efforts will focus on developing more efficient data generation strategies and exploring adaptive solutions to further enhance the performance and generalizability of CPSN for various super-resolution microscopy techniques.

**Supporting Information**

Supporting Information is available from the Wiley Online Library or from the author.

**ACKNOWLEDGMENTS**

This work was supported by the National Natural Science Foundation of China (62275125); National Major Scientific Research Instrument Development Project (62227818); The Fundamental Research Funds for the Central Universities (30922010313).

**Data Availability Statement**

The data that support the findings of this study are available from the corresponding author upon reasonable request.